\documentclass{ws-p8-50x6-00}

\input{psfig}

\newcommand{\peeppi}{$p(\vec e,e'\vec p\,)\pi^0$~}
\newcommand{\snp}{S_{0+}^{*}}

\newcommand{\sep}{S_{1+}^{*}}
\newcommand{\snpx}{S_{0+}}
\newcommand{\semx}{S_{1-}}
\newcommand{\sepx}{S_{1+}}

\newcommand{\eep}{E_{1+}}

\newcommand{\mep}{M_{1+}}

\begin{document}

\title{$\pi^0$  electroproduction in the $\Delta(1232)$  region at MAMI}

\author{H. SCHMIEDEN}

\address{Institut f\"{u}r Kernphysik, Johannes Gutenberg-Universit\"{a}t,
J.~J.~Becher-Weg 45,\\
D-55099 Mainz, Germany\\
E-mail: hs@kph.uni-mainz.de}

\maketitle

\abstracts{ 
            The extraction of the CMR from a \peeppi measurement
            is discussed. Preliminary results from further asymmetry measurements 
            with polarized and unpolarized electron beam indicate that the imaginary
            background is well under control in the MAID2000 parameterization,
            but not the real (Born-) amplitudes.
          }

\section{Introduction}\label{sec:introduction}

Similar to the ground state properties of the nucleon, 
the ratio of quadrupole to dipole strength in the $N \rightarrow \Delta(1232)$
transition since a long time was considered as one of the key observables
for the understanding of nucleon structure.
Several mechanisms had been proposed for the breaking of spherical symmetry
which is required for non-zero values for that ratio.

While the resonance structure of the nucleon has been revealed almost entirely 
by $\pi$-$N$  scattering, the $\Delta l = 2$  positive parity transitions
require electromagnetic excitations.
Due to the $\Delta(1232)$  decay branching ratio of $99.4$\,\% into the $\pi N$
channel, the electric (transverse) and Coulombic (longitudinal) quadrupole to
magnetic dipole ratios, EMR and CMR, respectively, are measured in pion photo-
and electroproduction.
These ratios are defined as 
$\mbox{EMR} = \Im m \{E_{1+}^{3/2}\} / \Im m \{M_{1+}^{3/2}\}$  and
$\mbox{CMR} = \Im m \{S_{1+}^{3/2}\} / \Im m \{M_{1+}^{3/2}\}$,
where the pion multipoles, $A^I_{l_{\pi}\pm}$, are characterized through their
magnetic, electric or longitudinal (scalar) nature, $A$,
the isospin, $I$, and the pion-nucleon relative angular momentum, $l_{\pi}$,
whose coupling with the nucleon spin
is indicated by $\pm$.                                                          
However, in pion production the $\Delta(1232)$  channel is intimately related to
non-resonant mechanisms, which hamper the extraction of the EMR and CMR.
The separation of resonant and non-resonant contributions, in principle, 
remains ambiguous \cite{Wilhelm96}.
What experimentally can be achieved is the isospin separation required for
the EMR and CMR.

However, a complete experiment with regard to a multipole decomposition seems
out of reach yet.
It requires the complete angular distributions of at least 7 or 11 independent
observables in single pion photo- and electroproduction, respectively.
The EMR and CMR thus can only be extracted using a truncated multipole basis or
model assumptions.
Therefore, it is desirable to measure the small quadrupole amplitudes in
different combinations with the unwanted, but unavoidable, 
non-resonant background.
In particular, the measurement of polarization observables is mandatory.
A first double-polarization experiment for the extraction of the CMR has been 
completed at MAMI.

\section{Results of the $p(\vec e,e'\vec p\,)\pi^0$  experiment}
\label{sec:Results}

High sensitivity to the CMR is provided in the $p(\vec e,e'\vec p\,)\pi^0$  reaction 
measured in parallel kinematics, both in the proton polarization component $P_x$  and
in the polarization ratio $P_x/P_z$  \cite{HS98}.
\begin{figure}[t]
\centerline{ \psfig{file=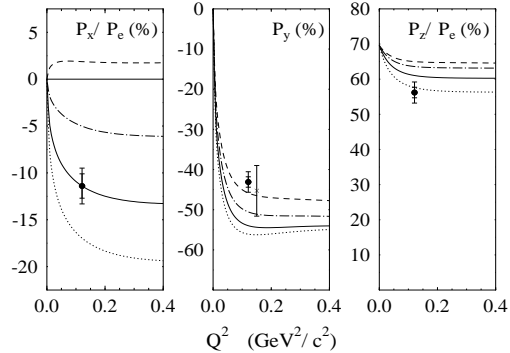,height=5.5cm,angle=0,silent=}
}
\caption{\label{fig:Pxyz}
          Results for the polarization components measured in the \peeppi reaction
          at $W=1232$\,MeV, $Q^2=0.12$\,(GeV/c)$^2$  and $\epsilon=0.71$
          in comparison with MAID2000
          calculations. The dashed, dot-dashed, full and dotted
          curves correspond to
          ${\rm CMR}=0$, $-3.2$, $-6.4$, $-9.6$\,\%, respectively.
          The MAMI data (full circles) are shown with statistical and
          systematical error.
          For the Bates $P_y$  (cross) only the statistical error is indicated,
          the value is rescaled in $\epsilon$  and, though measured at the same
          $Q^2$, slightly shifted for clarity. }
\end{figure}
\begin{figure}[t]
\begin{center}
\epsfxsize=18pc 
\epsfbox{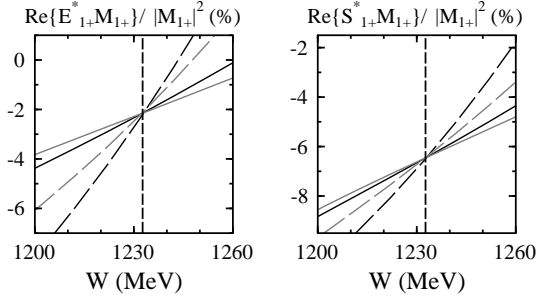} 
\caption{\label{fig:pi0-iso}
         $\Re{e}\{\eep^*\mep/|\mep|^2\}$  (left) and
         $\Re{e}\{\sep\mep/|\mep|^2\}$  (right) as a function of the
         invariant energy. The full and broken curves represent MAID2000
         calculations in the isospin 3/2 and the $p\pi^0$
         charge channel, respectively, for $Q^2=0.2$\,(GeV/c)$^2$  (black)
         and $1.0$\,(GeV/c)$^2$  (grey). The zero crossing of
         $\Re{e}\{M_{1+}^{3/2}\}$  is indicated by the vertical dashed line.
        }
\end{center}
\end{figure}
The axes are defined by
$ \hat{y} = \vec k_i \times \vec k_f / |\vec k_i \times \vec k_f|$, 
$ \hat{z} = \vec q / |\vec q\,|$  and
$ \hat{x} = \hat{y} \times \hat{z}$,
where $\vec k_i$  and $\vec k_f$  are the momenta of incoming and scattered
electron, respectively, and $\vec q = \vec k_i - \vec k_f$  denotes the momentum
transfer.
The experiment was carried out at MAMI with a 15\,$\mu$A electron beam of $P_e=75\,\%$
longitudinal polarization impinging on a 5\,cm long liquid hydrogen target.
The scattered electrons were detected in Spectrometer B of the 
3-Spectrometer-Setup \cite{Blomqvist98} in coincidence with the protons in
Spectrometer A.
This spectrometer was equipped with a focal plane proton polarimeter \cite{Pospischil01a}.
Due to the spin precession in the magnetic spectrometer and the helicity
flip of the electron beam, it was possible to determine all three proton polarization
components, $P_x$, $P_y$  and $P_z$, simultaneously.
The results \cite{Pospischil01}  are depicted in Figure\,\ref{fig:Pxyz}.
They were averaged over the experimental acceptance and projected to nominal parallel
kinematics ($W=1232$\,MeV, $Q^2=0.12$\,(GeV/c)$^2$, $\epsilon=0.71$, 
$\Theta_\pi^{\rm cm}=180^\circ$) using MAID2000 \cite{Drechsel99}.
A consistency relation among the reduced polarizations \cite{ST00} 
seems to be violated \cite{Pospischil01}. 
With the present statistical accuracy the origin of this remains unclear.

\begin{figure}[b]
\begin{center}
\epsfxsize=16pc 
\epsfbox{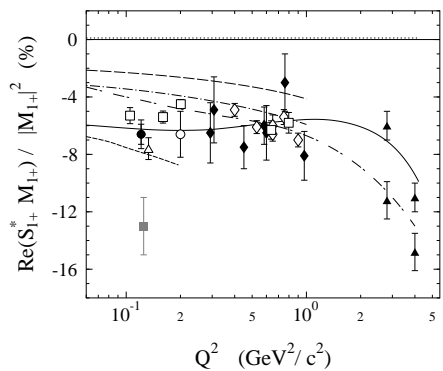} 
\caption{\label{fig:CMR}
          Results for $\Re{e}\{ S_{1+}^* M_{1+} \} / |M_{1+}|^2$
          extracted from different experiments.
          The pre 1980 data are indicated as full diamonds
          {\protect \cite{Siddle71,Alder72,Batzner74}}.
          The full grey square is from {\protect \cite{Kalleicher97}}.
          Full triangles represent the analyses of \protect\cite{Frolov99}.
          The result from the only double polarization measurement
          is indicated by the full circle with statistical (inner)
          and quadratic sum of statistical and systematical (outer) error
          \protect\cite{Pospischil01}.
          Preliminary data are also shown from Bates \protect\cite{Mertz99}
          (open triangle), TJNAF/CLAS \protect\cite{Burkert00} (open diamonds)
          ELSA \protect\cite{Gothe00a} (open squares) and MAMI
          \protect\cite{Elsner00,Suele01} (open circle).
          The curves are calculations within Heavy Baryon Chiral Perturbation
          Theory \protect\cite{Gellas99} (short dashed),
          the Chiral Quark Soliton Model \protect\cite{Silva00}
          SU(2) (dashed-dotted) and SU(3) (long dashed),
          the dynamical model of \protect\cite{SL00} (long-short dashed),
          and dressed (full) and bare (dotted) solutions of the dynamical
          model of \protect\cite{KY99}.                                              
         }
\end{center}
\end{figure}

In the MAID analysis CMR is extracted without truncation of the multipole
series.
Directly at the zero crossing of the real part of the $M_{1+}^{3/2}$ amplitude,
i.e. at $W=1232$\,MeV, the result obtained in the $p \pi^0$  channel is almost
identically with the isospin 3/2 channel.
This is demonstrated in Figure\,\ref{fig:pi0-iso}, where the relevant interferences
are shown as a function of $W$  in both channels.
Therefore, the CMR can be reliably extracted from the 
$p(\vec e,e'\vec p\,)\pi^0$  reaction,
without explicit isospin separation. 
In Figure\,\ref{fig:CMR} our result for $\Re{e}\{ S_{1+}^* M_{1+} \} / |M_{1+}|^2$
in the $p\pi^0$  channel is compared to previous results 
\cite{Siddle71,Alder72,Batzner74,Kalleicher97,Frolov99}
and the preliminary results of ongoing experiments
\cite{Mertz99,Burkert00,Gothe00a,Elsner00,Suele01}.  
The curves show the calculations of Ref.\,\cite{Gellas99,Silva00,SL00,KY99}.

\begin{figure}
\begin{center}
\epsfxsize=12pc 
\epsfbox{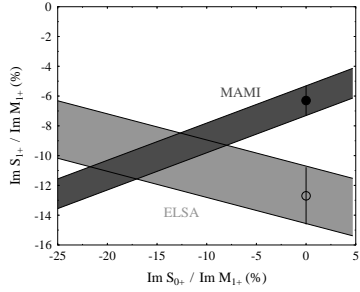} 
\caption{\label{fig:s0-s1}
         $\Im{m}\sepx/\Im{m}\mep$  as extracted as a function of
         $\Im{m}\snpx/\Im{m}\mep$  from measurements of $R_{LT}$-like
         structure functions with the pion in forward direction
         \protect\cite{Kalleicher97} (light shaded band) and backward direction
         \protect\cite{Pospischil01} (dark shaded band).
         The data points with errors correspond to the respective
         results at $\Im{m}\snpx=0$. 
        }
\end{center}
\end{figure}

The double-results at high $Q^2$  are due to different analyses of the
same data\,\cite{Frolov99}.
The model dependence of the analyses is reduced at smaller $Q^2$.
Here all results agree, except that of Ref.\,\cite{Kalleicher97} where,
in contrast to the recoil polarization measurement, 
the $\pi^0$  was detected in forward direction and all multipoles other than
$\mep$  and $\sepx$  were neglected.
The effect of an additionally non-zero $\Im{m}\snpx/\Im{m}\mep$  on the extracted 
$\Im{m}\sepx/\Im{m}\mep$  is visualized in Figure\,\ref{fig:s0-s1}.
The extracted  $\Im{m}\sepx/\Im{m}\mep$  is plotted as a function of 
$\Im{m}\snpx/\Im{m}\mep$.
Within MAID2000 the imaginary part of $\snpx$  has a strong impact on the extraction 
of the CMR, too.
$\Im{m}\snpx$  can be measured via the forward-backward asymmetry of $LT$-type
structure functions \cite{HS98a,Kelly99}.

\section{Forward and backward LT-asymmetry in $p(e,e'p)\pi^0$}
\label{sec:Forward}

Due to a different angular weight of $\Re{e}\sep\mep$  and $\Re{e}\snp\mep$,
their relative strength 
can be extracted from the $\Theta_\pi^{\rm cm}$  angular distribution.
This requires the LT structure function or the related asymmetry
$\rho_{LT} = \frac{d\sigma^L - d\sigma^R}{d\sigma^L + d\sigma^R}$
to be measured over a large angular range \cite{HS98a}.
$d\sigma^L$  and $d\sigma^R$  denote the differential cross sections left and right 
of $\vec q$.

At the squared 4-momentum transfer of $Q^2=0.2$\,(GeV/c)$^2$  of the MAMI experiment,
the forward ($\Theta_\pi^{\rm cm} = 20^\circ$) and 
backward ($\Theta_\pi^{\rm cm} = 160^\circ$) kinematics require very different
settings of the proton spectrometer both in angle and momentum.
They are summarized in Table\,\ref{tab:kinematics}.
\begin{table}[htbp]
\begin{center}
\caption{\label{tab:kinematics} Kinematics of the $p(e,e'p)\pi^0$  experiment.
         $\Theta_p^{\rm lab}$  and $p_p^{\rm lab}$  are the proton laboratory angles and 
         momenta, respectively, and $\Theta_{pq}^{\rm lab}$  denotes the relative angle 
         between recoiling proton and momentum transfer. For all settings the angle
         of the momentum transfer is $26.9^\circ$ relative to the incoming electron beam.
        } \vskip 0.0 cm
\begin{tabular}{ccccc}
\hline
 kinematics  &  $\Theta_\pi^{\rm cm}$  &  $\Theta_{pq}^{\rm lab}$  &  $p_p^{\rm lab}$  (MeV/c)  &  $\Theta_p^{\rm lab}$ \\ \hline
\hline
 $b$  (right) & $160^\circ$   & $6.1^\circ$    & $741.7$  & $33.0^\circ$ \\
 $b$  (left)  &               &                &          & $20.9^\circ$ \\
\hline
 $f$ (right)  &  $20^\circ$   & $17.2^\circ$   & $265.0$  & $44.2^\circ$ \\
 $f$ (left)   &               &                &          &  $9.7^\circ$ \\
\hline
\end{tabular}
\end{center}
\end{table}
A 10\,$\mu$A (unpolarized) beam was used on a LH$_2$  target with a diameter of
only 1\,cm, in order to minimize the energy loss of the low energy protons of the
forward settings (c.f. Table\,\ref{tab:kinematics}).
The scattered electrons were detected in Spectrometer A and the protons in Spectrometer B.

\begin{figure}
\begin{center}
\epsfxsize=16pc 
\epsfbox{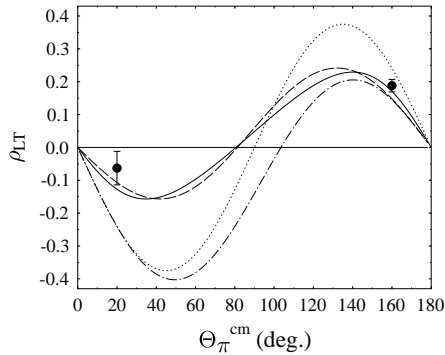} 
\caption{\label{fig:rho_LT}
         Left-right asymmetries for the $p(e,e'p)\pi^0$  cross section
         around $W=1232$\,MeV, $Q^2=0.2$\,(GeV/c)$^2$ and $\epsilon=0.6$.
         Forward and backward proton kinematics,
         $\Theta_{\pi}^\mathrm{cm} = 160^\circ$  and
         $\Theta_{\pi}^\mathrm{cm} = 20^\circ$, respectively
         \protect\cite{Elsner00,Suele01}, were measured at identical
         electron kinematics. 
         The full curve is from MAID2000 while the other curves are obtained in
         a truncated multipole basis (only $\mep$, $\sepx$  and $\snpx$  partial waves) with 
         $\Im{m}{\sepx}/\Im{m}{\mep}=-6.6\,\%$, $\Im{m}{\snpx}/\Im{m}{\mep}=+6.0\,\%$ 
         (the MAID standard values --- dashed),
         $-12.5\,\%$, $0\,\%$  (dotted), and
         $-10.0\,\%$, $-14\,\%$  (dashed-dotted).
        }
\end{center}
\end{figure}

In Figure\,\ref{fig:rho_LT} the preliminary results \cite{Elsner00,Suele01} are shown 
in comparison with calculations.
The full curve corresponds to the full MAID calculation which has the standard ratios
$\Im{m}{\sepx}/\Im{m}{\mep}=-6.6\,\%$  and $\Im{m}{\snpx}/\Im{m}{\mep}=+6.0\,\%$.
The other curves are due to a truncated multipole basis where only 
$\mep$, $\sepx$  and $\snpx$  partial waves are included.
The differences between the full and dashed curves indicate the quality of 
this approximation.
From the result of Ref.\,\cite{Kalleicher97} the dotted curve would be expected,
while the dashed-dotted line corresponds to the overlap region in Figure\,\ref{fig:s0-s1}.
The new data do not yet indicate any discrepancy with the
standard MAID parameterization concerning the 
$\Im{m}{\sepx}/\Im{m}{\mep}$  and $\Im{m}{\snpx}/\Im{m}{\mep}$  ratios.
Thus, the imaginary parts of the relevant amplitudes seem well under control.

\section{Fifth structure function in $p(\vec e,e'p)\pi^0$}
\label{sec:Fifth}

In contrast to the imaginary parts, the real background is more problematic.
This is indicated by the discrepancy between MAID calculation and data for $P_y$
(c.f. Figure\,\ref{fig:Pxyz}, middle).
$P_y$  is related to the imaginary part of a longitudinal-transverse interference.
In parallel kinematics
the leading terms in s and p wave approximation read \cite{HS98}:
\begin{equation}
P_y \propto  \Im{m}\{(S_{0+} - 4 S_{1+} + S_{1-})^* \mep \}.
\label{eq:P_y_sp}
\end{equation}
This is very similar to the fifth structure function in $p(\vec e,e'p)\pi^0$:
\begin{equation}
R_{LT'} \propto P_e \sin{\Theta_\pi^{\rm cm}}\,\Im{m}
                \{ (\snpx +6 \cos\Theta_\pi^{\rm cm} \sepx)^* \mep \}.
\label{eq:R_LTs}                                               
\end{equation}

\begin{figure}
\begin{center}
\centerline{ \psfig{file=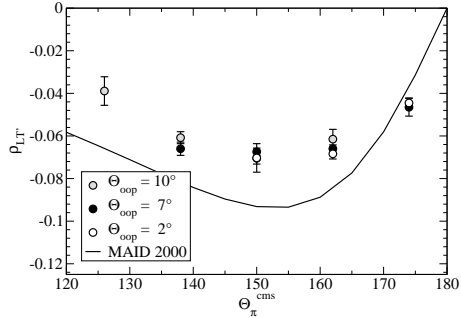,height=6.0cm,angle=-90,silent=}}
\caption{\label{fig:rho_LTs}
         Cross section asymmetry $\rho_{LT'}$  for the 
         $p(\vec e,e'p)\pi^0$  reaction
         around $W=1232$\,MeV and $Q^2=0.2$\,(GeV/c)$^2$,
         measured at $\Phi=90^\circ$  (out of plane)
         \protect\cite{Bartsch-Dr}.
         White, grey and black dots indicate $2^\circ$, $7^\circ$  and $10^\circ$
         out of plane settings of the proton spectrometer, respectively.
         The curve is due to the full MAID2000 calculation.
        }
\end{center}
\end{figure}

Therefore, we measured the helicity asymmetry 
$\rho_{LT'} = \frac{d\sigma^+ - d\sigma^-}{d\sigma^+ + d\sigma^-}$  with regard to
the flip of the electron beam helicity between $+$ and $-$.
This required detection of the recoiling proton in Spectrometer
B with its $10^\circ$  out of plane capability. 
The preliminary result is shown in Figure\,\ref{fig:rho_LTs}.
Obviously, the MAID parameterization, very similarly to $P_y$,  also overestimates 
the magnitude of $\rho_{LT'}$.
It is therefore concluded that this is not caused by the $\semx$  amplitude, 
which only contributes to $P_y$  (Eq.\,\ref{eq:P_y_sp}), 
but is rather related to the real (Born) background in the imaginary parts of
the interferences of Eqs.\,\ref{eq:P_y_sp} and \ref{eq:R_LTs}.
Such a real background to first order does not affect the extraction of the
CMR from $P_x$  at the resonance energy.

\section{Summary}

From a first \peeppi double polarization experiment 
${\rm CMR} = (-6.4 \pm 0.7_{\rm stat} \pm 0.8_{\rm syst})\,\%$  has been deduced
within the MAID2000 parameterization.
Imaginary background amplitudes investigated by forward and backward $\rho_{LT}$  measurements 
are in agreement with MAID. 
Contrary, MAID consistently overestimates the magnitude of $P_y$  as well as of $\rho_{LT'}$,
which was measured with out of plane detection of the recoiling protons.
This seems due to real amplitudes which, at the resonance energy, to first order don't affect 
the extraction of the CMR. 
Its model uncertainty is thus estimated of the order of $1\,\%$  absolute.

\section*{Acknowledgements}
I thank all my colleagues from Basel, Ljubljana, Rutgers and Mainz,
in particular the students P. Bartsch, D. Elsner, S. Gr\"ozinger, Th. Pospischil, and
A. S\"ule.
This work was supported by the Deutsche Forschungsgemeinschaft (SFB 443).


\begin{thebibliography}{99}

\bibitem{Wilhelm96}     P. Wilhelm et al., Phys. Rev. \textbf{C 54}, 1423 (1996) 
\bibitem{HS98}          H. Schmieden, Eur. Phys. J. \textbf{A 1}, 427 (1998)     
\bibitem{Blomqvist98}   K.I. Blomqvist et al., 
                        Nucl. Instr. Methods {\bf A 403}, 263 (1998)
\bibitem{Pospischil01a} Th. Pospischil et al.,
                        submitted to Nucl. Instr. Meth. \textbf{A}
                        and nucl-ex/0010007 
\bibitem{Pospischil01}  Th. Pospischil et al., 
                        Phys. Rev. Lett. \textbf{86}, 2959 (2001)
\bibitem{Drechsel99}    D. Drechsel, O. Hanstein, S.S. Kamalov and L. Tiator,
                        Nucl. Phys. {\bf A 645}, 145 (1999)  and
                        http://www.kph.uni-mainz.de/MAID/maid2000/
\bibitem{ST00}          H. Schmieden and L. Tiator, 
                        Eur. Phys. J. \textbf{A 8}, 15 (2000)
\bibitem{Siddle71}      R. Siddle et al., Nucl. Phys. \textbf{B 35}, 93 (1971)
\bibitem{Alder72}       J.C. Alder et al., Nucl. Phys. \textbf{B 46}, 573 (1972)
\bibitem{Batzner74}     K. B{\"a}tzner et al., 
                        Nucl. Phys. \textbf{B 76}, 1 (1974) 
\bibitem{Kalleicher97}  F. Kalleicher et al., Z. Phys. \textbf{A 359}, 201 (1997)
\bibitem{Frolov99}      V.V. Frolov et al., 
                        Phys. Rev. Lett. \textbf{82}, 45 (1999)
\bibitem{Mertz99}       C. Mertz et al., nucl-ex/9902012 
\bibitem{Burkert00}     V.D. Burkert,  
                        proceedings of Few Body Problems in Physics,
                        Taipei (2000),  hep-ph/0007297
\bibitem{Gothe00a}      R. Gothe,
                        proceedings of NSTAR2000, Newport News (2000)
\bibitem{Elsner00}      D. Elsner, Diploma thesis, Mainz (2000) (unpublished)
\bibitem{Suele01}       A. S\"ule, Diploma thesis, Mainz (2001) (unpublished)
\bibitem{Gellas99}      G.C. Gellas et al.,
                        Phys. Rev. \textbf{D 60}, 054022 (1999)
\bibitem{Silva00}       A. Silva et al.,
                        Nucl. Phys. \textbf{A 675}, 637 (2000)
\bibitem{SL00}          T. Sato and T.-S.H. Lee,
                        Phys. Rev. \textbf{C 63}, 055201 (2001)
\bibitem{KY99}          S.S. Kamalov and S.N. Yang,
                        Phys. Rev. Lett. \textbf{83}, 4494 (1999)
\bibitem{HS98a}         H. Schmieden et al., approved MAMI proposal A1-3/98
\bibitem{Kelly99}       J.J. Kelly, Phys. Rev. \textbf{C 60}, 054611 (1999)
\bibitem{DT92}          D. Drechsel and L. Tiator,
                        J. Phys. G: Nucl. Part. Phys. \textbf{18}, 449 (1992)
\bibitem{Bartsch-Dr}    P. Bartsch, doctoral thesis, Mainz, in preparation

\end{thebibliography}
\end{document}